%% file: prl-H-tautau.tex
\documentclass[aps,prl,twocolumn,showpacs,groupedaddress]{revtex4}  
\usepackage{graphicx}  
\usepackage{dcolumn}   
\usepackage{bm}        
\usepackage{amssymb}   

\newcommand{\etana}{\mbox{$e\tau_h$}}
\newcommand{\mtana}{\mbox{$\mu\tau_h$}}
\newcommand{\emana}{\mbox{$e\mu$}}
\newcommand{\ttype}{\mbox{$\tau$-type}}

\newcommand{\hs}{\ensuremath{\mathit{\Phi}}}
\newcommand{\ppbar}{\mbox{$p\overline{p}$}}
\newcommand{\pbs}{\mbox{$\rm{pb}^{-1}$}}

\newcommand{\dzero}{\mbox{D\O}}
\newcommand{\ttbar}{\mbox{$t \overline{t} $}}
\newcommand{\bbbar}{\mbox{$b \overline{b} $}}
\newcommand{\tanb}{\mbox{$\tan \beta$}}
\newcommand{\Gev}{${\rm{GeV}}$}
\newcommand{\Tev}{${\rm{TeV}}$}
\newcommand{\Pet}{\mbox{\ensuremath{\not \!\! P_T}}}
\newcommand{\met}{\mbox{\ensuremath{\not \!\! E_T}}}
\newcommand{\metx}{\mbox{\ensuremath{\not \!\! E_x}}}
\newcommand{\mety}{\mbox{\ensuremath{\not \!\! E_y}}}
\newcommand{\Gc}{${\rm{GeV}}$}

\def \mvis {\ensuremath{ M_{\rm{vis}} }}
\def \lep  {\ensuremath{ \ell }}

\begin{document}

\hspace{5.2in} \mbox{Fermilab-Pub-06/092-E}

\title{Search for Neutral Higgs Bosons Decaying to Tau Pairs \\ 
in \mbox{$p\overline{p}$} Collisions at $\sqrt{s}$ = 1.96 TeV}
\input list_of_authors_r2.tex  

\date{May 2, 2006}

\begin{abstract}
A search for the production of neutral Higgs bosons \hs\ decaying into $\tau^+\tau^-$ final states in \ppbar\ collisions at a center-of-mass energy of 1.96~TeV is presented. The data, corresponding to an integrated luminosity of approximately 325~\pbs, were collected by the \dzero\ experiment at the Fermilab Tevatron Collider. Since no excess compared to the expectation from standard model processes is found, limits on the production cross section times branching ratio are set. 
The results are combined with those obtained from the D\O\ search for $\hs b (\bar{b})\rightarrow\bbbar b (\bar{b})$ and are interpreted in the minimal supersymmetric standard model.

\end{abstract}

\maketitle 

Final states leading to high-mass tau lepton pairs can arise from various physics processes beyond the standard model (SM) including the production of neutral Higgs bosons (generally denoted as \hs ). Higgs bosons are an essential ingredient of electroweak symmetry breaking in the SM, but so far remain unobserved experimentally. A search for Higgs bosons decaying to tau leptons is of particular interest in models with more than one Higgs doublet, where production rates for $p\bar{p}\rightarrow\hs\rightarrow\tau\tau$ can potentially be large enough for an observation at the Fermilab Tevatron Collider. For instance, the minimal supersymmetric standard model (MSSM) \cite{MSSM} contains two complex Higgs doublets, leading to two neutral CP-even ($h,H$), one CP-odd ($A$), and a pair of charged ($H^{\pm}$) Higgs bosons. At tree level, the Higgs sector of the MSSM is fully specified by two parameters, generally chosen to be $M_A$, the mass of the CP-odd Higgs boson, and \tanb, the ratio of the vacuum expectation values of the two Higgs doublets. At large \tanb, the coupling of the neutral Higgs bosons to down-type quarks and charged leptons is strongly enhanced, leading to sizeable cross sections and increased decay rates to the third generation tau lepton and bottom quark. MSSM scenarios with large \tanb\ are of considerable interest since they can provide a viable dark matter candidate \cite{battaglia}.

Searches for neutral MSSM Higgs bosons have been conducted at LEP \cite{LEP-limit} and at the Tevatron \cite{dzero-bb,CDF-tautau}. In this Letter a search for $\hs \to \tau \tau$ decays is presented. At least one of the tau leptons is required to decay leptonically, leading to final states containing $e  \tau_h$, $\mu  \tau_h$ and $e \mu$, where $\tau_h$ represents a hadronically decaying tau lepton. 

The data were collected at the Fermilab Tevatron Collider between September 2002 and August 2004 at $\sqrt{s} = 1.96$~TeV and correspond to integrated luminosities of 328~\pbs, 299~\pbs, and 348~\pbs\ for the $e \tau_h$, $\mu \tau_h$ and $e \mu$ final states, respectively. Final states with two electrons or two muons have a small signal-to-background ratio due to the small branching fraction and the large background from $Z/\gamma^*$ production, and are therefore not considered.

A thorough description of the \dzero\ detector can be found in Ref.~\cite{d0det}. Briefly, the detector consists of a magnetic central tracking system surrounded by a liquid-argon and uranium calorimeter and a toroidal muon spectrometer. The central tracking system comprises  a silicon microstrip tracker (SMT) and a central fiber tracker (CFT), both located within a 2~T magnetic field provided by a superconducting solenoidal magnet. The SMT and CFT designs were optimized to provide precise tracking and vertexing capabilities over the pseudorapidity range $|\eta|<2.5$, where $\eta = -{\rm ln}(\tan(\theta/2))$ and $\theta$ is the polar angle with respect to the proton beam. The calorimeter is divided into a central section covering $|\eta| \lesssim 1.1$, and two end calorimeters (EC) that extend coverage to $|\eta|\approx 4.2$. A muon system, at $|\eta|<2$, consists of a layer of tracking detectors and scintillation trigger counters in front of 1.8~T toroids, followed by two similar layers after the toroids. The luminosity is measured by detecting inelastic \ppbar\ scattering processes in plastic scintillator arrays located in front of the EC cryostats, covering $2.7 < |\eta| < 4.4$.

The \etana\ and the \mtana\ analyses rely on single electron and single muon triggers, respectively, while the \emana\ analysis uses dilepton triggers. 
Signal and SM processes are modeled using the {\sc pythia}~6.202~\cite{pythia} Monte Carlo (MC) generator, followed by a {\sc geant}-based~\cite{geant} simulation of the D\O\ detector geometry. All background processes, apart from QCD multijet production, are normalized using cross sections calculated  at next-to-leading order (NLO) and next-to-NLO (for $Z$ boson, $W$ boson, and Drell-Yan production) based on the CTEQ5~\cite{cteq} parton distribution functions (PDF). 

The normalization and shape of background contributions from QCD multijet production, where jets are misidentified as leptons, are estimated from the data by using like-sign $e$ and $\tau_h$ candidate events (\etana\ analysis) or by selecting  background samples by inverting lepton identification criteria (\mtana\ and $e \mu$ analyses). These samples are normalized to the data at an early stage of the selection in a region of phase space dominated by multijet production.

Isolated electrons are reconstructed based on their characteristic energy deposition in the calorimeter, including the transverse and longitudinal shower profile. In addition, a track must point to the energy deposition in the calorimeter, and the track momentum and calorimeter energy must be consistent. Further rejection against background from photons and jets is achieved by using a likelihood discriminant, which is exploiting characteristic calorimeter and tracking information. Muons are selected using tracks in the central tracking detector in combination with patterns of hits in the muon detector. Muons are required to be isolated in both the calorimeter and the tracker. Reconstruction efficiencies for both leptons are measured using data.

A hadronically decaying tau lepton is characterized by a narrow isolated jet with low track multiplicity. The tau reconstruction is either seeded by calorimeter energy clusters or tracks~\cite{d0-z-tautau}. Three \ttype s are distinguished:

\begin{itemize}
\item \ttype\ 1: a single track with energy deposition in the hadronic calorimeter (1-prong, $\pi^\pm$-like);
\item \ttype\ 2: a single track with energy deposition in the hadronic and the electromagnetic calorimeter (1-prong, $\rho^\pm$-like);
\item \ttype\ 3: two or three tracks with an invariant mass below 1.1 or 1.7~\Gev, respectively (3-prong).
\end{itemize}

A set of neural networks, one for each \ttype, has been developed based on further discriminating variables. The neural networks were used elsewhere for a cross section measurement of the process $Z/\gamma^*\rightarrow\tau\tau$~\cite{d0-z-tautau}. The input variables exploit the differences between hadronically decaying tau leptons and jets in the longitudinal and transverse shower shape as well as differences in the isolation in the calorimeter and the tracker. The training of the neural networks is performed using multijet events from data as the background sample and tau MC events as signal, resulting in a network output close to one for tau candidates and close to zero for background. Both training samples cover the kinematic region of interest for this analysis. For \ttype s 1 and 2, hadronic tau candidates are required to have a neural network output greater than $0.9$. Due to the larger background contamination, this cut value is tightened to $0.95$ for \ttype~3. 

Electrons and muons can be misidentified as one-prong hadronic tau decays. Hadronically decaying tau leptons deposit a significant fraction of their energy in the hadronic part of the calorimeter. To reject electrons, the ratio between the transverse energy in the hadronic calorimeter and the transverse momentum of the tau track is required to be larger than 0.4. With a smaller rate, background from muons occurs in \ttype s~1 and 2 in the $\mtana$ analysis. This background is suppressed by rejecting tau candidates to which a muon can be matched.

The signal is characterized by two leptons, missing transverse energy $\met$, and little jet activity. It would stand out as an enhancement above the background from SM processes in the visible mass 
\begin{equation}
\mvis  =  \sqrt{(P_{\tau_1}\ + \ P_{\tau_2} \ + \ \Pet)^2}, 
\end{equation}
calculated using the four vectors of the visible tau decay products $P_{\tau_{1,2}}$ and of the missing momentum $\Pet = (\met,\metx,\mety, 0)$. $\metx$ and $\mety$ indicate the components of $\met$. For the optimization of the signal selection, only the high mass region is used, which is defined as $\mvis > 120$~\Gev\ in the \etana\ and \mtana\ analyses and as $\mvis > 110$~\Gev\ in the \emana\ analysis. 

In the \etana\ and \mtana\ analyses, an isolated lepton ($e, \mu$) and an isolated hadronic tau with transverse momenta above 14~\Gc\ and 20~\Gc\ respectively are required. In addition to the irreducible background from $Z/\gamma^*\rightarrow\tau\tau$ production, a $W\rightarrow\lep\nu$ decay can be misidentified as a high-mass di-tau event if it is produced in association with an energetic jet that is misidentified as a hadronic tau decay. In these events, a strongly boosted $W$ boson recoils against the jet, and the mass of the $W$ boson can be reconstructed in the following approximation $M_W^{e/\mu} =  \sqrt{2 \ E^\nu \ E^{e/\mu} \ (1-\cos\Delta\phi)}$, where the azimuthal angle $\Delta\phi$ is between the lepton and $\met$,  and $E^\nu=\met\cdot E^{e/\mu}/E_T^{e/\mu}$. $M_W^{e/\mu}$ is required to be less than $20$~GeV.

In the \emana\ analysis, two isolated leptons each with \mbox{$p_T >$ 14~\Gc} are required. The dominant background contributions after the lepton selection come from the irreducible $Z/\gamma^* \rightarrow \tau \tau$ process, followed by $WW$, $WZ$, $\ttbar$, $W \rightarrow \ell \nu$, and multijet events. In this analysis the multijet background is suppressed by requiring $\met > 14$~\Gc. Background from $W$+jet events can be reduced using the transverse mass $M_T^{e/\mu}=\sqrt{2 \ p_T^{e/\mu} \ \met \ (1-\cos\Delta\phi)}$ by requiring that either $M_T^e<10$~GeV or $M_T^\mu<10$~GeV. Furthermore the minimum angle between the leptons and the \met\ vector, $\min[\Delta\phi(e,\met),\Delta\phi(\mu,\met)]$, has to be be smaller than $0.3$. Finally, contributions from \ttbar\ background are suppressed by a cut on the scalar sum of the transverse momenta of all jets in the event $H_T < 70$~\Gc.

\begin{table}
\caption{Numbers of events observed in data and expected for background and the efficiency for a signal with \mbox{$M_\hs=150$~GeV} for the three analysis channels, with statistical and systematic uncertainties added in quadrature. \label{t:evt_numbers}}
\begin{ruledtabular}
\begin{tabular}{lr@{\hspace{.3ex}$\pm$\hspace{-3.7ex}}lr@{\hspace{.3ex}$\pm$\hspace{-3.7ex}}lr@{\hspace{.3ex}$\pm$\hspace{-3.7ex}}l}
Analysis         & \multicolumn{2}{c}{$e \tau_h$ }    & \multicolumn{2}{c}{$\mu \tau_h$}  & \multicolumn{2}{c}{  $e \mu$} \\
\hline
Data             &   \multicolumn{2}{c}{  337   }      &  \multicolumn{2}{c}{575  }        & \multicolumn{2}{c}{  41} \\  \hline
QCD                              &  144 & 19   &  62   & 7     &  2.1  & 0.4 \\ 
$Z / \gamma^* \to \tau \tau$     &  130 & 17   & 492   & 53    &  39   & 5   \\
$Z / \gamma^* \to ee, \mu \mu$   &   12 & 2    &   5   & 1     &  0.6  & 0.1 \\
$W \to e \nu, \mu \nu, \tau \nu$ &    9 & 1    &  14   & 2     &  0.3  & 0.2 \\
Di-boson                         &  0.4 & 0.1  &  3.1  & 0.3   &  1.0  & 0.1 \\ 
$\ttbar$                         &  0.3 & 0.1  &  1.2  & 0.2   &  0.06 & 0.02\\ \hline
Total expected                   &  296 & 38   & 576   & 62    &  44   & 5   \\ \hline
Efficiency \%                    &  3.6 & 0.4  &  8.6  & 0.8   &  4.3  & 0.5 \\
\end{tabular}\\[5mm]
\end{ruledtabular}
\end{table}

\begin{figure}[t!!!]
\begin{center}
\includegraphics[height=73mm]{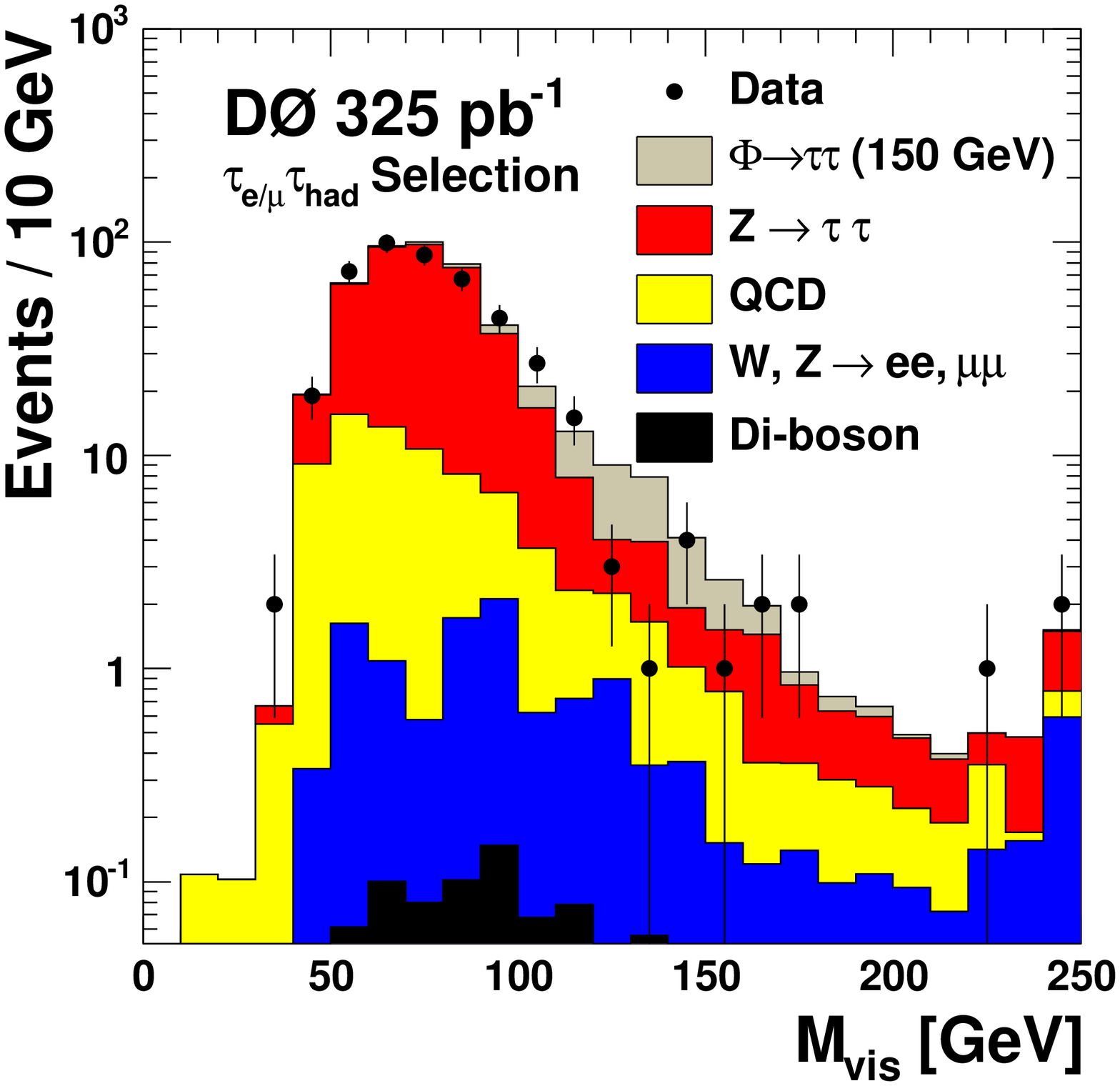}
\includegraphics[height=73mm]{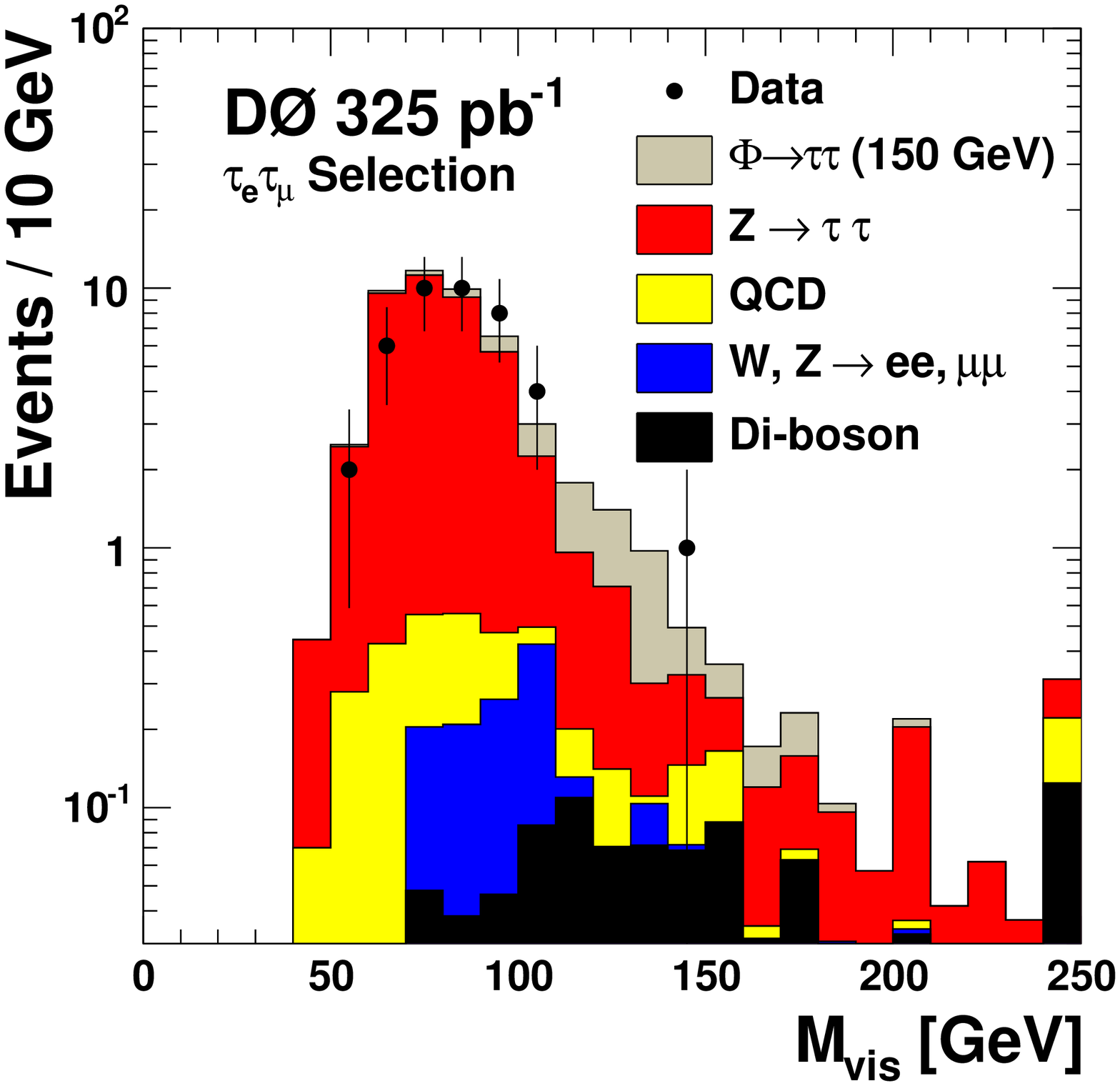}
\caption[]{The distribution of the visible mass $\mvis$ for the two final states involving hadronic tau decays and for the \emana\ final state. The Higgs signal is normalized to the cross section excluded by this analysis. The upper distribution shows the subsample with the largest signal-to-background ratio ($M_W^{e , \mu} <6$~\Gev). The highest bin includes the overflow, the indicated luminosity represents the average of the three final states.}
\label{f:vis-mass}
\end{center}
\end{figure}

The numbers of events observed in the data and those expected from the various SM processes show good agreement, as can be seen in Table~\ref{t:evt_numbers} and Fig.~\ref{f:vis-mass}.
The estimate of the expected numbers of background and signal events depends on numerous measurements that introduce a systematic uncertainty: integrated luminosity (6.5\%), trigger efficiency (1\%--4\%), lepton identification and reconstruction efficiencies (2\%--5\%), jet and tau energy calibration (2\%--6\%), PDF uncertainty (3\%--4\%), and modeling of multijet background (2\%--9\%). All except the last one are correlated between the three final states.

\begin{figure}[t!!!]
\begin{center}
\includegraphics[height=73mm]{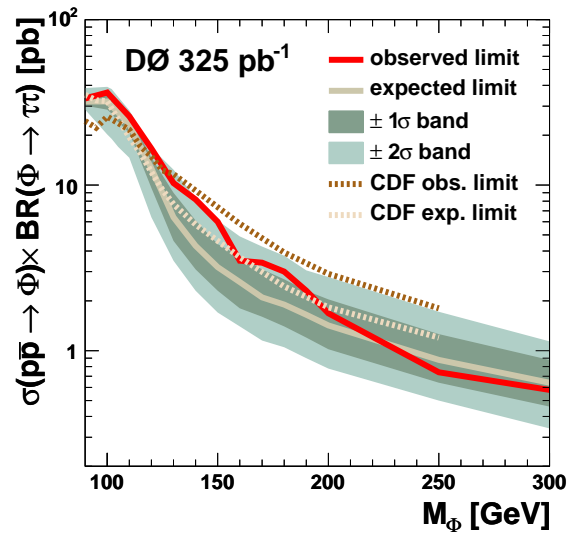}
\caption[]{The observed and expected 95\% CL limits on the cross section times branching ratio for $\hs \to \tau \tau$ production as a function of $M_\hs$ assuming a narrow width of the Higgs boson. The error bands include systematic and statistical uncertainties. CDF curves are taken from~\cite{CDF-tautau}, where data corresponding to an integrated luminosity of 310~\pbs\ is used.}
\label{f:xs-limit}
\end{center}
\end{figure}

The efficiencies for a Higgs boson signal are found to vary between 1.6\%, 4.0\%, and 1.2\% for $M_{\hs}$ = 100~\Gev\ and 8.3\%, 13.6\%, and 9.3\% for $M_{\hs}$ = 300~\Gev\ for the $e \tau_h$, $\mu \tau_h$, and $e \mu$ analyses respectively. Since no significant evidence for the production of neutral Higgs bosons with decays $\hs \to \tau \tau$ is observed, upper limits on the production cross section times branching ratio are extracted as a function of $M_\hs$. In order to maximize the sensitivity (expected limit), the event samples of the \etana\ and \mtana\ analyses are split into subsamples according to different signal-to-background ratios: The subsamples are separated by \ttype\ and by $M_W$ ($M_W^{e , \mu} <6$~\Gev, \mbox{$6 < M_W^{e , \mu} <20$}~\Gev).
Furthermore the differences in shape between signal and background are exploited by using the information of the full mass spectrum of $\mvis$ in the limit calculation. Both the expected and the observed limits on the cross section times branching ratio at the 95\% confidence level (CL), calculated using the modified frequentist approach \cite{cls}, are presented in Fig.~\ref{f:xs-limit}.

In the MSSM, the masses and couplings of the Higgs bosons depend, in addition to \tanb\ and $M_A$, on the SUSY parameters through radiative corrections. In a constrained model, where unification of the SU(2) and U(1) gaugino masses is assumed, the most relevant parameters are the mixing parameter $X_t$, the Higgs mass parameter $\mu$, the gaugino mass term $M_2$, the gluino mass $m_g$, and a common scalar mass $M_{\rm SUSY}$.  
Limits on $\tan \beta$ as a function of $M_A$ are derived for two scenarios assuming a CP-conserving Higgs sector: the so-called $m_h^{\rm max}$ scenario (with the parameters  $M_{\rm SUSY}$= 1~\Tev, $X_t$ = 2~\Tev, $M_2$ = 0.2~\Tev, $\mu$ = $\pm$0.2~\Tev, and $m_g$ = 0.8~\Tev) and the no-mixing scenario (with the parameters  \mbox{$M_{\rm SUSY}$= 2~\Tev}, $X_t$ = 0, $M_2$ = 0.2~\Tev, \mbox{$\mu$ = $\pm$0.2~\Tev}, and $m_g$ = 1.6~\Tev)~\cite{mssm-benchmark}. The production cross sections, widths, and branching ratios for the Higgs bosons are calculated over the mass range from 90 to 300~\Gev\ using the {\sc feynhiggs} program \cite{feynhiggs}, where the complete set of one-loop corrections and all known two-loop corrections are incorporated. The contributions of SUSY particles in the loop of the gluon fusion process are taken into account, as well as mass- and \tanb -dependent decay widths. In the region of large \tanb, the $A$ boson is nearly degenerate in mass with either the $h$ or the $H$ boson, and their production cross sections are added.

\begin{figure}
\begin{center}
\includegraphics[height=73mm]{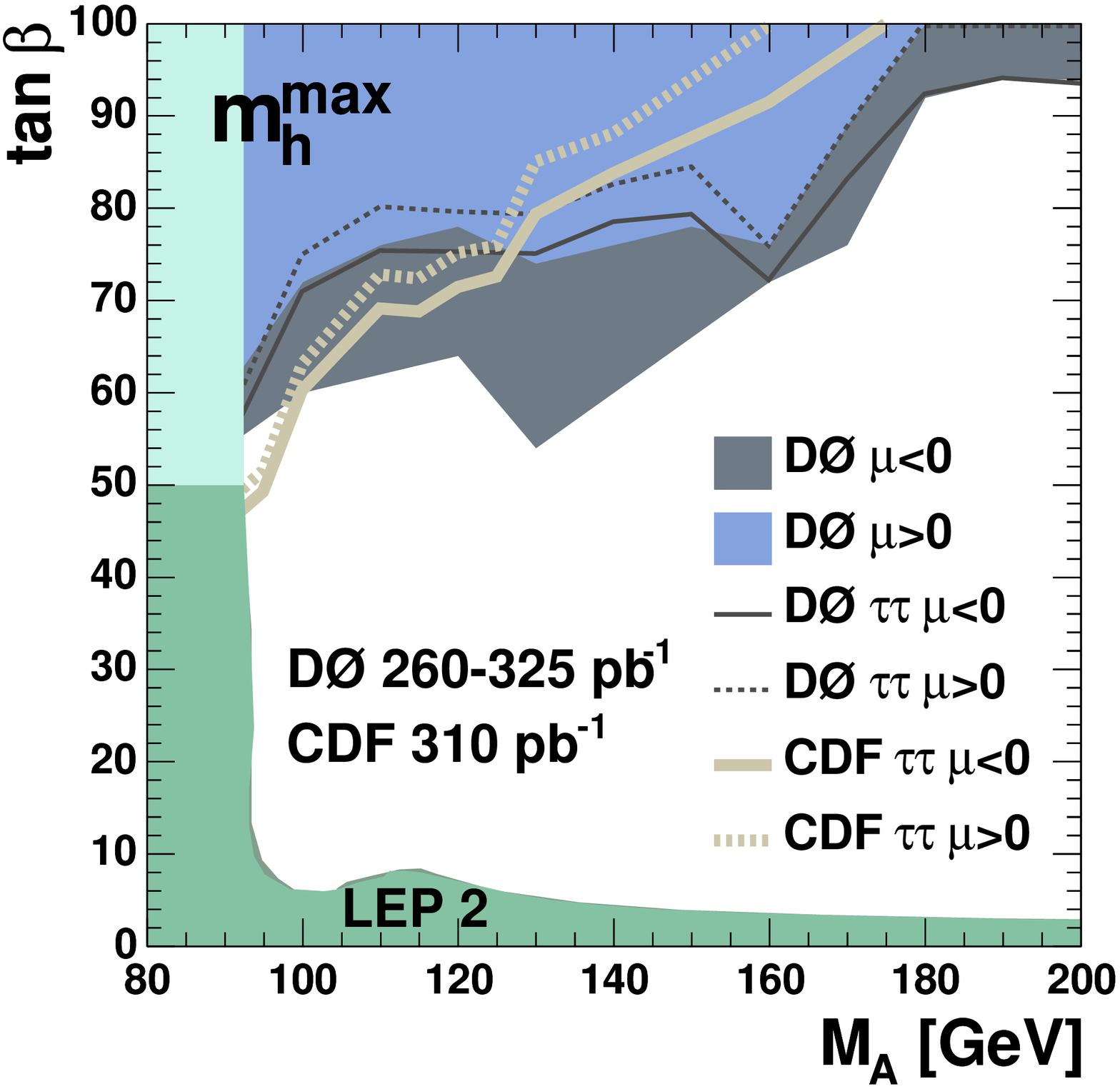}
\includegraphics[height=73mm]{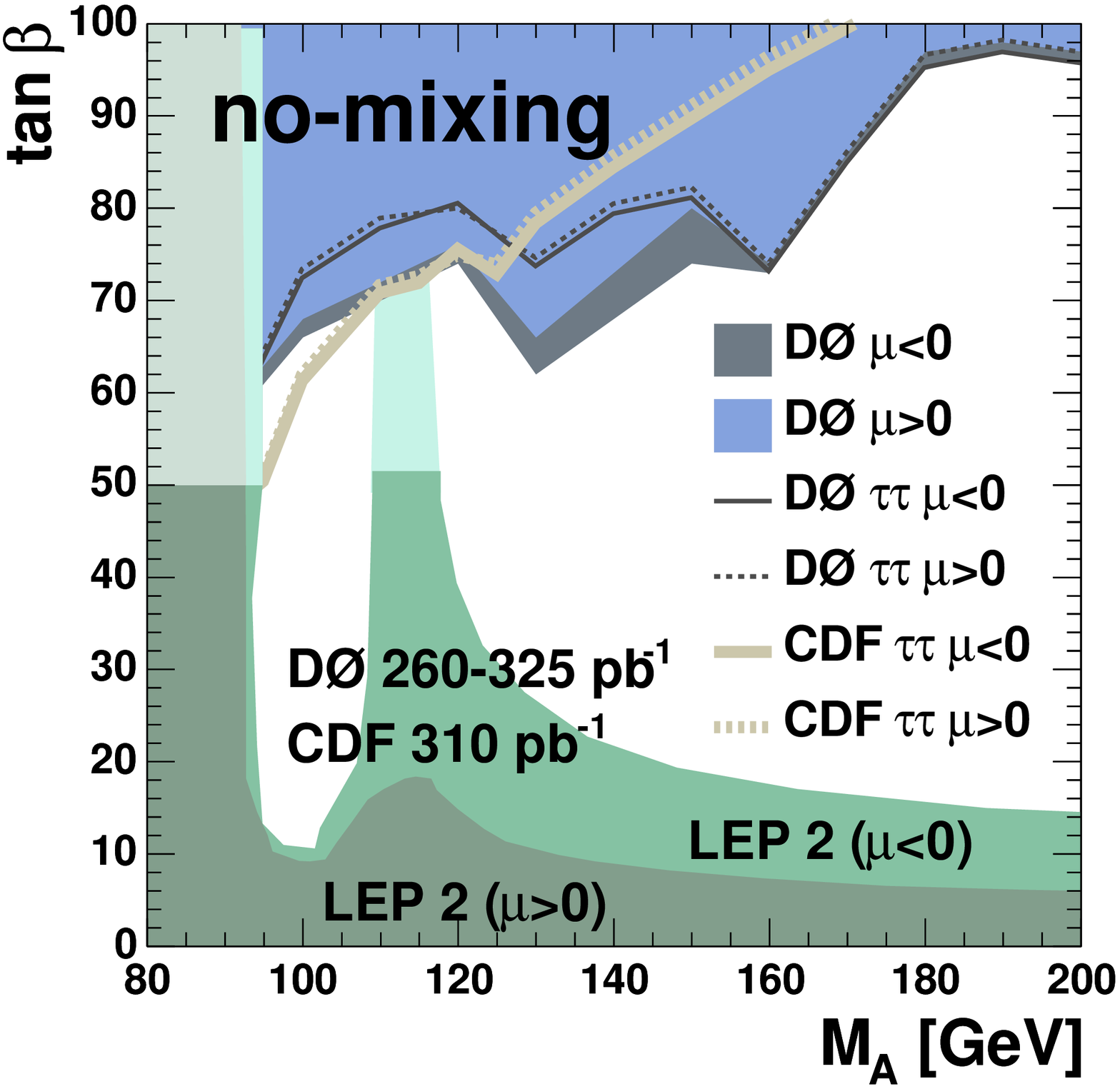}
\caption[]{Region, which is excluded at 95\% CL, in the ($M_A, \tanb$) plane for the $m_h^{\rm max}$ and the no-mixing scenario for $\mu =+0.2$~TeV and $\mu = -0.2$~TeV \mbox{($m_{\rm top}=172.7$~GeV)}. The results obtained in the present analysis (labeled as D\O\ $\tau\tau$) are combined with those obtained in the $\hs b (\bar{b})\rightarrow\bbbar b (\bar{b})$ search~\cite{dzero-bb}. The LEP limits~\cite{LEP-limit} have been extrapolated for $\tanb > 50$.} 
\label{f:tanb-excl}
\end{center}
\end{figure}

Fig.~\ref{f:tanb-excl} shows the \dzero\ results obtained in the present analysis in combination with those obtained in the  $\hs b (\bar{b})\rightarrow\bbbar b (\bar{b})$ search~\cite{dzero-bb}, which are re-interpreted in the MSSM scenarios used in this Letter. 
The combined result currently represents the most stringent limit on the production of neutral MSSM Higgs bosons at hadron colliders. 

\input{acknowledgement_paragraph_r2}
\end{document}

%% file: list_of_authors_r2.tex
%
\author{                                                                      
V.M.~Abazov,$^{36}$                                                           
B.~Abbott,$^{76}$                                                             
M.~Abolins,$^{66}$                                                            
B.S.~Acharya,$^{29}$                                                          
M.~Adams,$^{52}$                                                              
T.~Adams,$^{50}$                                                              
M.~Agelou,$^{18}$                                                             
J.-L.~Agram,$^{19}$                                                           
S.H.~Ahn,$^{31}$                                                              
M.~Ahsan,$^{60}$
% ...
\\                                                                            
\vskip 0.30cm                                                                 
\centerline{(D\O\ Collaboration)}                                             
\vskip 0.30cm                                                                 
}                                                                             
\affiliation{                                                                 
\centerline{$^{1}$Universidad de Buenos Aires, Buenos Aires, Argentina}       
\centerline{$^{2}$LAFEX, Centro Brasileiro de Pesquisas F{\'\i}sicas,         
                  Rio de Janeiro, Brazil}                                     
\centerline{$^{3}$Universidade do Estado do Rio de Janeiro,                   
                  Rio de Janeiro, Brazil}                                     
\centerline{$^{4}$Instituto de F\'{\i}sica Te\'orica, Universidade            
                  Estadual Paulista, S\~ao Paulo, Brazil}                     
\centerline{$^{5}$University of Alberta, Edmonton, Alberta, Canada,           
                  Simon Fraser University, Burnaby, British Columbia, Canada,}
\centerline{York University, Toronto, Ontario, Canada, and                    
                  McGill University, Montreal, Quebec, Canada}                
\centerline{$^{6}$Institute of High Energy Physics, Beijing,                  
                  People's Republic of China}                                 
\centerline{$^{7}$University of Science and Technology of China, Hefei,       
                  People's Republic of China}                                 
\centerline{$^{8}$Universidad de los Andes, Bogot\'{a}, Colombia}             
\centerline{$^{9}$Center for Particle Physics, Charles University,            
                  Prague, Czech Republic}                                     
\centerline{$^{10}$Czech Technical University, Prague, Czech Republic}        
\centerline{$^{11}$Center for Particle Physics, Institute of Physics,         
                   Academy of Sciences of the Czech Republic,                 
                   Prague, Czech Republic}                                    
\centerline{$^{12}$Universidad San Francisco de Quito, Quito, Ecuador}        
\centerline{$^{13}$Laboratoire de Physique Corpusculaire, IN2P3-CNRS,         
                   Universit\'e Blaise Pascal, Clermont-Ferrand, France}      
\centerline{$^{14}$Laboratoire de Physique Subatomique et de Cosmologie,      
                   IN2P3-CNRS, Universite de Grenoble 1, Grenoble, France}    
\centerline{$^{15}$CPPM, IN2P3-CNRS, Universit\'e de la M\'editerran\'ee,     
                   Marseille, France}                                         
\centerline{$^{16}$IN2P3-CNRS, Laboratoire de l'Acc\'el\'erateur              
                   Lin\'eaire, Orsay, France}                                 
\centerline{$^{17}$LPNHE, IN2P3-CNRS, Universit\'es Paris VI and VII,         
                   Paris, France}                                             
\centerline{$^{18}$DAPNIA/Service de Physique des Particules, CEA, Saclay,    
                   France}                                                    
\centerline{$^{19}$IReS, IN2P3-CNRS, Universit\'e Louis Pasteur, Strasbourg,  
                    France, and Universit\'e de Haute Alsace,                 
                    Mulhouse, France}                                         
\centerline{$^{20}$Institut de Physique Nucl\'eaire de Lyon, IN2P3-CNRS,      
                   Universit\'e Claude Bernard, Villeurbanne, France}         
\centerline{$^{21}$III. Physikalisches Institut A, RWTH Aachen,               
                   Aachen, Germany}                                           
\centerline{$^{22}$Physikalisches Institut, Universit{\"a}t Bonn,             
                   Bonn, Germany}                                             
\centerline{$^{23}$Physikalisches Institut, Universit{\"a}t Freiburg,         
                   Freiburg, Germany}                                         
\centerline{$^{24}$Institut f{\"u}r Physik, Universit{\"a}t Mainz,            
                   Mainz, Germany}                                            
\centerline{$^{25}$Ludwig-Maximilians-Universit{\"a}t M{\"u}nchen,            
                   M{\"u}nchen, Germany}                                      
\centerline{$^{26}$Fachbereich Physik, University of Wuppertal,               
                   Wuppertal, Germany}                                        
\centerline{$^{27}$Panjab University, Chandigarh, India}                      
\centerline{$^{28}$Delhi University, Delhi, India}                            
\centerline{$^{29}$Tata Institute of Fundamental Research, Mumbai, India}     
\centerline{$^{30}$University College Dublin, Dublin, Ireland}                
\centerline{$^{31}$Korea Detector Laboratory, Korea University,               
                   Seoul, Korea}                                              
\centerline{$^{32}$SungKyunKwan University, Suwon, Korea}                     
\centerline{$^{33}$CINVESTAV, Mexico City, Mexico}                            
\centerline{$^{34}$FOM-Institute NIKHEF and University of                     
                   Amsterdam/NIKHEF, Amsterdam, The Netherlands}              
\centerline{$^{35}$Radboud University Nijmegen/NIKHEF, Nijmegen, The          
                  Netherlands}                                                
\centerline{$^{36}$Joint Institute for Nuclear Research, Dubna, Russia}       
\centerline{$^{37}$Institute for Theoretical and Experimental Physics,        
                   Moscow, Russia}                                            
\centerline{$^{38}$Moscow State University, Moscow, Russia}                   
\centerline{$^{39}$Institute for High Energy Physics, Protvino, Russia}       
\centerline{$^{40}$Petersburg Nuclear Physics Institute,                      
                   St. Petersburg, Russia}                                    
\centerline{$^{41}$Lund University, Lund, Sweden, Royal Institute of          
                   Technology and Stockholm University, Stockholm,            
                   Sweden, and}                                               
\centerline{Uppsala University, Uppsala, Sweden}                              
\centerline{$^{42}$Physik Institut der Universit{\"a}t Z{\"u}rich,            
                   Z{\"u}rich, Switzerland}                                   
\centerline{$^{43}$Lancaster University, Lancaster, United Kingdom}           
\centerline{$^{44}$Imperial College, London, United Kingdom}                  
\centerline{$^{45}$University of Manchester, Manchester, United Kingdom}      
\centerline{$^{46}$University of Arizona, Tucson, Arizona 85721, USA}         
\centerline{$^{47}$Lawrence Berkeley National Laboratory and University of    
                   California, Berkeley, California 94720, USA}               
\centerline{$^{48}$California State University, Fresno, California 93740, USA}
\centerline{$^{49}$University of California, Riverside, California 92521, USA}
\centerline{$^{50}$Florida State University, Tallahassee, Florida 32306, USA} 
\centerline{$^{51}$Fermi National Accelerator Laboratory,                     
            Batavia, Illinois 60510, USA}                                     
\centerline{$^{52}$University of Illinois at Chicago,                         
            Chicago, Illinois 60607, USA}                                     
\centerline{$^{53}$Northern Illinois University, DeKalb, Illinois 60115, USA} 
\centerline{$^{54}$Northwestern University, Evanston, Illinois 60208, USA}    
\centerline{$^{55}$Indiana University, Bloomington, Indiana 47405, USA}       
\centerline{$^{56}$University of Notre Dame, Notre Dame, Indiana 46556, USA}  
\centerline{$^{57}$Purdue University Calumet, Hammond, Indiana 46323, USA}    
\centerline{$^{58}$Iowa State University, Ames, Iowa 50011, USA}              
\centerline{$^{59}$University of Kansas, Lawrence, Kansas 66045, USA}         
\centerline{$^{60}$Kansas State University, Manhattan, Kansas 66506, USA}     
\centerline{$^{61}$Louisiana Tech University, Ruston, Louisiana 71272, USA}   
\centerline{$^{62}$University of Maryland, College Park, Maryland 20742, USA} 
\centerline{$^{63}$Boston University, Boston, Massachusetts 02215, USA}       
\centerline{$^{64}$Northeastern University, Boston, Massachusetts 02115, USA} 
\centerline{$^{65}$University of Michigan, Ann Arbor, Michigan 48109, USA}    
\centerline{$^{66}$Michigan State University,                                 
            East Lansing, Michigan 48824, USA}                                
\centerline{$^{67}$University of Mississippi,                                 
            University, Mississippi 38677, USA}                               
\centerline{$^{68}$University of Nebraska, Lincoln, Nebraska 68588, USA}      
\centerline{$^{69}$Princeton University, Princeton, New Jersey 08544, USA}    
\centerline{$^{70}$State University of New York, Buffalo, New York 14260, USA}
\centerline{$^{71}$Columbia University, New York, New York 10027, USA}        
\centerline{$^{72}$University of Rochester, Rochester, New York 14627, USA}   
\centerline{$^{73}$State University of New York,                              
            Stony Brook, New York 11794, USA}                                 
\centerline{$^{74}$Brookhaven National Laboratory, Upton, New York 11973, USA}
\centerline{$^{75}$Langston University, Langston, Oklahoma 73050, USA}        
\centerline{$^{76}$University of Oklahoma, Norman, Oklahoma 73019, USA}       
\centerline{$^{77}$Oklahoma State University, Stillwater, Oklahoma 74078, USA}
\centerline{$^{78}$Brown University, Providence, Rhode Island 02912, USA}     
\centerline{$^{79}$University of Texas, Arlington, Texas 76019, USA}          
\centerline{$^{80}$Southern Methodist University, Dallas, Texas 75275, USA}   
\centerline{$^{81}$Rice University, Houston, Texas 77005, USA}                
\centerline{$^{82}$University of Virginia, Charlottesville,                   
            Virginia 22901, USA}                                              
\centerline{$^{83}$University of Washington, Seattle, Washington 98195, USA}  
}                                                                             

%% file: acknowledgement_paragraph_r2.tex
%
We thank S. Heinemeyer for useful discussions.
We thank the staffs at Fermilab and collaborating institutions, 
and acknowledge support from the 
DOE and NSF (USA);
CEA and CNRS/IN2P3 (France);
FASI, Rosatom and RFBR (Russia);
CAPES, CNPq, FAPERJ, FAPESP and FUNDUNESP (Brazil);
DAE and DST (India);
Colciencias (Colombia);
CONACyT (Mexico);
KRF and KOSEF (Korea);
CONICET and UBACyT (Argentina);
FOM (The Netherlands);
PPARC (United Kingdom);
MSMT (Czech Republic);
CRC Program, CFI, NSERC and WestGrid Project (Canada);
BMBF and DFG (Germany);
SFI (Ireland);
The Swedish Research Council (Sweden);
Research Corporation;
Alexander von Humboldt Foundation;
and the Marie Curie Program.